\documentclass[aps,prb,twocolumn,amsmath,amssymb]{revtex4-1}

\usepackage{epsfig}
\usepackage{graphicx}
\usepackage{amsfonts}
\usepackage{amssymb}
\usepackage{amsmath}
\usepackage{bm}

\renewcommand{\vec}[1]{{\mathbf #1}}

\bibpunct{(}{)}{;}{s}{,}{,}

\begin{document}

\title{Pseudospin Transfer Torques in Semiconductor Electron Bilayers}

\author{Youngseok Kim}
\affiliation{Department of Electrical and Computer Engineering, University of Illinois, Urbana, Il, 61801}
\altaffiliation{Micro and Nanotechnology Laboratory, University of Illinois, Urbana, Il 61801}

\author{A. H. MacDonald}
\affiliation{Department of Physics, University of Texas at Austin,  Austin, Texas 78712}

\author{Matthew J. Gilbert}
\affiliation{Department of Electrical and Computer Engineering, University of Illinois, Urbana, Il, 61801}
\altaffiliation{Micro and Nanotechnology Laboratory, University of Illinois, Urbana, Il 61801}
\altaffiliation{Beckman Institute, University of Illinois, Urbana, Il 61801}
\altaffiliation{Institute for Condensed Matter Theory, University of Illinois, Urbana, Il 61801}

\date{\today}

\begin{abstract}
We use self-consistent quantum transport theory to investigate the influence of electron-electron interactions 
on interlayer transport in semiconductor electron bilayers in the absence of an external magnetic field. 
We conclude that, even though spontaneous pseudospin order does not occur at zero field, interaction-enhanced quasiparticle tunneling amplitudes and pseudospin transfer torques do alter tunneling I-V characteristics, and 
can lead to time-dependent response to a {\em dc} bias voltage.  
\end{abstract}


\maketitle

\section{Introduction} 
In this paper we address interlayer transport in separately contacted nanometer length scale semiconductor bilayers, with a view toward the identification of possible interaction-induced collective transport effects. When interaction effects are neglected a nanoscale conductor always reaches a steady state\cite{Datta2000} in which current increases smoothly with bias voltage. 
Nanoscale transport theory has at its heart the evaluation of the density-matrix of non-interacting electrons in contact with two or more reservoirs whose chemical potentials differ. This problem is efficiently solved using Green's function techniques, for example by using the non-equilibrium Green's function (NEGF) method\cite{Datta2000}. Real electrons interact, of course, and the free-fermion degrees of freedom which appear in this type of theory should always be thought of as Fermi liquid theory\cite{Shankar1994, Pines1966, Giuliani2005} quasiparticles. The effective single-particle Hamiltonian therefore depends on 
the microscopic configuration of the system. In practice the quasiparticle Hamiltonian is often\cite{Ullrich2002, Ventra2001} calculated from a self-consistent mean-field theory like Kohn-Sham density-functional-theory (DFT). DFT, spin-density functional theory, current-density functional theory, Hartree theory, and Hartree-Fock theory all function as useful fermion self-consistent-field theories. Since a bias voltage changes the system density matrix, it inevitably changes the 
quasiparticle Hamiltonian. Determination of the steady state density matrix therefore requires a self-consistent calculation. 

Self-consistency is included routinely in NEGF simulations\cite{Gilbert2004, Gilbert2005, Wang2004} of transport at the Hartree theory level in nanoscale semiconductor systems and in simulations\cite{Sergueev2007, Wu2005, Waldron2006, Toher2007, Rocha2005} of molecular transport. The collective behavior captured by self-consistency can sometimes change the current-voltage relationship in a qualitative way, leading to an I-V curve that is not smooth or even to circumstances in which there is no steady state response to a time-independent bias voltage. Some of the most useful and interesting examples of this type of effect occur in ferromagnetic metal spintronics. The quasiparticle Hamiltonian is a ferromagnetic metal has a large spin-splitting term which lowers the energy of quasiparticles whose spins are aligned with the magnetization (majority-spin quasiparticles) relative to those quasiparticles whose spins are aligned opposite to the magnetization (minority-spin quasiparticles). When current flows in a ferromagnetic metal the magnetization direction is altered.\cite{Nenez2006}  The resulting change in the quasiparticle Hamiltonian\cite{Nenez2006} is responsible for the rich variety of so-called \textit{spin-transfer torque} collective transport effects which occur in magnetic metals and semiconductors. These spin-transfer torques\cite{Haney2008, Haney2007} are often understood macroscopically\cite{Comment1} as the \textit{reaction} counterpart of  the torques which act on the quasiparticle spins that carry current through a non-collinear ferromagnet. Spin-transfer torques\cite{Haney2008} can lead to discontinuous I-V curves and to oscillatory\cite{Rippard2006} or chaotic response to a time-independent bias voltage. Similar phenomena\cite{Rossi2005} occur in semiconductor bilayers for certain ranges of external magnetic field over which the ground state has spontaneous\cite{Wen1992, Moon1995, Eisenstein2004} interlayer phase coherence or (equivalently) exciton condensation. Indeed when bilayer exciton condensate ordered states are viewed as pseudospin ferromagnets, the spectacular transport anomalies\cite{Eisenstein2004, Spielman2000, Gunawan2004, Wiersma2004, Kellogg2002} they exhibit have much in common with those of ferromagnetic metals.

In this article we consider semiconductor bilayers at zero magnetic field, using a pseudospin language\cite{MacDonald2001, Jungwirth2001} in which top layer electrons are said to have pseudospin up ($\vert\uparrow\rangle$) and bottom layer electrons are said to have pseudospin down ($\vert\downarrow\rangle$). Although inter-layer transport in semiconductor bilayers has been studied extensively in the strong field quantum Hall regime, work on the zero field limit has been relatively sparse and has focused on studies of interlayer drag,\cite{Misra2008, Eisenstein1991, Eisenstein19912, Zheng1993} counterflow,\cite{Su2008} and on speculations about possible broken symmetry states.\cite{MacDonald1988, Ruden1991} We concur with the consensus view that pseudospin ferromagnetism is not expected in conduction band two-dimensional electron systems\cite{Comment2, Senatore2003} except possibly\cite{Stern2000} at extremely low carrier densities. There are nevertheless pseudospin-dependent interaction contributions to the quasiparticle Hamiltonian. When a current flows the pseudospin orientation of transport electrons is altered, just as in metal spintronics, and some of the same phenomena can occur. The resulting change in the quasiparticle Hamiltonian is responsible for a current-induced pseudospin-torque which alters the state of non-transport electrons well away from the Fermi energy. Indeed although the spin-splitting field appears spontaneously in ferromagnetic metals, spintronics phenomena usually depend on an interplay between the spontaneous exchange field and effective magnetic fields due to magnetic-dipole interactions, spin-orbit coupling, and external magnetic fields. In semiconductor bilayers the pseudospin external field is due to single-particle interlayer tunneling amplitude and is typically of the same order\cite{Swierkowski1997} as the interaction 
contribution to the pseudospin-splitting field.

The paper is organized in the following manner. We start in Section II with a qualitative discussion which explains the role played in transport by exchange correlation enhancement of the quasiparticle interlayer tunneling amplitude. We also briefly outline the methods and the approximations that we have employed in the illustrative numerical calculations which are the main subject of this paper. In Section III, we describe the results of self-consistent quantum transport calculation for an ideal disorder-free system with two GaAs quantum wells separated by a small AlGaAs barrier. For balanced double-layer electron systems, the inter-layer current has a resonant peak near zero-bias.  Because the current is a non-monotonic function of inter-layer bias voltage, it is not immediately obvious that it is possible to drive a current that is large enough to induce exchange related instabilities. The calculations in Section III demonstrate that interactions alter the conductance peak near zero bias which occurs in weakly disordered systems in the absence of a strong magnetic field, and that exchange interactions can destabilize transport steady states. Finally, in Section IV, we summarize our results.

\begin{figure}[t!]   
  \centering
    \includegraphics[width=0.45\textwidth]{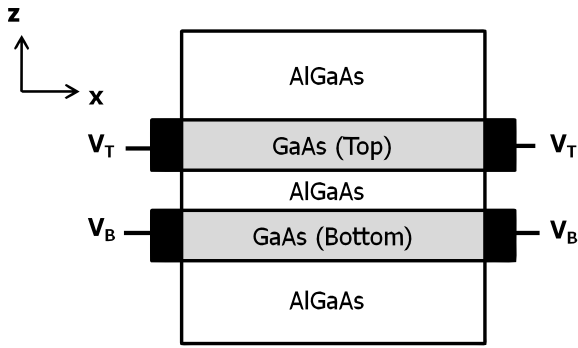}
  \caption{Device cross-section in $x, z$ direction. Top and bottom Al\(_{0.9}\)Ga\(_{0.1}\)As barrier thickness is 60 nm, GaAs quantum well thickness is 15 nm, and Al\(_{0.9}\)Ga\(_{0.1}\)As barrier thickness is 1 nm. A channel is $1.2~\mu m$ long and $7.5~\mu m$ wide. An assumed electron density in 2 dimensional electron gas (2DEG) is $2.0\times10^{10}cm^{-2}$.} \label{FIG:Device}
\end{figure}  

\section{Bilayer Pseudospin Torques} 
\subsection{Interacting Bilayer Model}  
We begin this section by briefly describing the approximations that we use to model semiconductor bilayer tunneling I-V characteristics. We consider an Al\(_{0.9}\)Ga\(_{0.1}\)As/GaAs bilayer heterostructure with $60$ nm top and bottom Al\(_{0.9}\)Ga\(_{0.1}\)As barriers which act to isolated the coupled quantum wells from electrostatic gates, as illustrated in 
Fig.~\ref{FIG:Device}.  The 
bilayer consists of two 15nm deep GaAs quantum wells with an assumed 2DEG electron density of \(2.0\times 10^{10}cm^{-2}\)
separated by a 1nm Al\(_{0.9}\)Ga\(_{0.1}\)As barrier. The quantum wells are  $1.2\mu  m$ long and \(7.5\mu m\) wide with the splitting between the symmetric and antisymmetric states set to a small value \(\Delta _{SAS}=2t=2\mu eV\). We define the z-axis as the growth direction, the x-axis as the longitudinal (transport) direction, 
and the y-axis as the direction across the transport channel as shown schematically in Fig. (\ref{FIG:Device}).  
We connect ideal contacts attached to the inputs and outputs of both layers.  The 
contacts inject and extract current and enter into the Hamiltonian via appropriate self-energy terms.\cite{Datta2000} 

We construct the system Hamiltonian from a model with a 
single band effective mass Hamiltonian for top and bottom layers 
and a phenomenological single-particle inter-layer tunneling term:
\begin{equation}
\label{Eq:Hdef}
\bold{H}=\begin{bmatrix}H_{TL} & 0 \\ 0 &H_{BL}\end{bmatrix} + \sum_{\mu = x,y,z} \hat{\mu} \cdot \vec{\Delta} \otimes \sigma_\mu .
\end{equation}
The first term on the right hand side of Eq. (\ref{Eq:Hdef}) is the single-particle non-interacting term 
while the second term is a mean-field interaction term.
In the second term in right hand side, $\sigma_\mu$ represents the Pauli spin matrices in each of the three spatial directions $\mu=x,y,z$, $\otimes$ represents the Kronecker product, and $\bold{\Delta}$ is a pseudospin effective magnetic field which will be discussed in more detail later in this section. To explore interaction physics in bilayer transport qualitatively, we use a local density approximation in which the interaction contribution to the quasiparticle Hamiltonian is proportional to the pseudospin-magnetization at each point in space. If we take the top layer as the pseudospin up state (\(\vert\uparrow\rangle\)) and the bottom layer as the pseudospin down state (\(\vert\downarrow\rangle\)), 
the single particle interlayer tunneling term 
contributes a pseudospin effective field \(\bold{H}\) with magnitude $t=\Delta_{SAS}/2$ and direction \(\hat{x}\). In real spin ferromagnetic systems, interactions between spin-polarized electrons lead to an effective magnetic field 
in the direction of spin-polarization.  Bilayers with pseudospin-polarization due to tunneling have a similar interaction contribution to the quasiparticle Hamiltonian.  Including both single-particle and many-body interaction contributions, the pseudospin effective field $\mathbf{\Delta}$ term in the quasiparticle Hamiltonian is,\cite{Gilbert2010, MacDonald2001}
\begin{equation}
\label{Eq:deltadef}
\bold{\Delta}=(t + U \bold{m}_{ps}^x) \, \hat{x}+ U \bold{m}_{ps}^y \, \hat{y} 
\end{equation}
where the pseudospin-magnetization \(m_{ps}\) is defined by 
\begin{equation}
\label{Eq:psdef}
\bold{m_{ps}}=\frac{1}{2}\operatorname{Tr}[\rho_{ps}\bm{\tau}].
\end{equation}
In Eq. (\ref{Eq:psdef}), $\bm{\tau}=\sigma_x,\sigma_y,\sigma_z$ is the vector of Pauli spin matrices, and $\rho_{ps}$ is the $2\times2$ Hermitian pseudospin density matrix which we define as,
\begin{equation}
\rho_{ps}=\begin{bmatrix} \rho_{\uparrow\uparrow}&\rho_{\uparrow\downarrow}\\\rho_{\downarrow\uparrow}&\rho_{\downarrow\downarrow} \end{bmatrix}.
\end{equation}
The diagonal terms of pseudospin density matrix (\(\rho_{\uparrow\uparrow}, \rho_{\downarrow\downarrow}\)) are the electron densities of top and bottom layers. In Eq. (\ref{Eq:deltadef}), we have dropped the exchange potential associated with the $\hat{z}$ component of pseudospin because it is dominated by the electric potential difference between layers induced by the inter-layer bias voltage (See below). 
From the definition of Eq. (\ref{Eq:Hdef}), the pseudospin-magnetization of $\hat{x},\hat{y},\hat{z}$ direction is defined 
in terms of the density matrix as
\begin{align}
&\langle m_{ps}^x\rangle=\frac{1}{2}(\rho_{\uparrow\downarrow}+ \rho_{\downarrow\uparrow}), \\
&\langle m_{ps}^y\rangle=\frac{1}{2}(-i\rho_{\uparrow\downarrow}+i \rho_{\downarrow\uparrow}), \\
&\langle m_{ps}^z\rangle=\frac{1}{2}(\rho_{\uparrow\uparrow}-\rho_{\downarrow\downarrow}).
\end{align}
As a result, we may express the system Hamiltonian in terms of pseudospin field contributions,
\begin{equation}
\bold{H}=\begin{bmatrix}H_{TL}+\Delta_z&\Delta_x-i\Delta_y \\ \Delta_x+i\Delta_y&H_{BL}-\Delta_z\end{bmatrix}.
\end{equation}
Here $\Delta_z$ is the electric potential difference between the two layers which we evaluate in a 
Hartree approximation, disregarding its exchange contribution.
The planar pseudospin angle which figures prominently in the discussion below is defined by 
\begin{equation}
\label{Eq:psfielddef}
\phi_{ps}=\tan^{-1} \left( \frac{\langle m_{ps}^y\rangle}{\langle m_{ps}^x\rangle} \right).
\end{equation}
This angle corresponds physically to the phase difference between electrons in the two layers.  

\begin{figure}[t!]  
  \centering
    \includegraphics[width=0.3\textwidth]{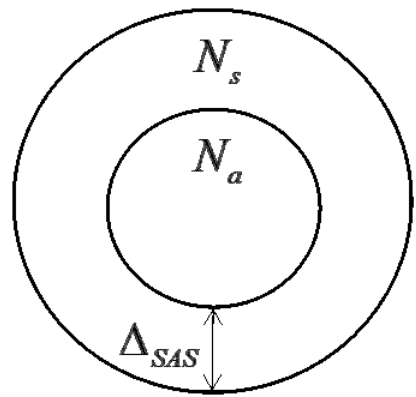}
  \caption{Schematic of the Fermi surfaces of the bilayer system containing populations of symmetric and antisymmetric state separated in energy by a single particle tunneling term, $\Delta_{SAS}$.}\label{FIG:Fermisurface}
\end{figure}   

\subsection{Enhanced Interlayer Tunneling} 
With the quasiparticle Hamiltonian for our semiconductor bilayer defined, we now address the 
strength of the interlayer interactions present in the system.
The interaction parameter $U$ in Eq. (\ref{Eq:deltadef}) is chosen so that the local density approximation for interlayer exchange reproduces a prescribed value for the interaction enhancement of the interlayer tunneling amplitude.  In equilibrium the 
pseudospin magnetization will be oriented in the $\hat{x}$ direction and the quasiparticle will have either 
symmetric or antisymmetric bilayer states with pseudospins in the $\hat{x}$ and $-\hat{x}$ 
directions respectively.  The majority and minority pseudospin states 
differ in energy at a given momentum by $2 t_{eff}$ where 
\begin{equation}
\label{Eq:teff}
t_{eff} = t + U \, \frac{N_s - N_a}{2} .
\end{equation}
The population difference between symmetric and antisymmetric differences may be evaluated from the 
differences in their Fermi radii illustrated in Fig.~\ref{FIG:Fermisurface}:
\begin{equation}
\label{Eq:nsna}
\frac{N_s - N_a}{2} = \nu_0 \, t_{eff}.
\end{equation}
where $\nu_0$ is the density-of-states of a single layer.
Combining Eq. (\ref{Eq:teff}) and Eq. (\ref{Eq:nsna}) we can relate $U$ to
$S$, the interaction enhancement factor for the interlayer tunneling amplitude:
\begin{equation}
\label{Eq:teff2}
t_{eff}=\frac{t}{1- U\nu_0} \equiv S \, t.
\end{equation}
The physics of $S$ is similar to that responsible for the interaction enhancement of the Pauli susceptibility in metals.  According to microscopic theory\cite{Swierkowski1997} a typical value for $S$ is around 2. 
We choose to use $S$ rather than $U$ as a parameter in our calculations and therefore set
%
\begin{equation}
U = \frac{1-S^{-1}}{\nu_0} \simeq \frac{1}{2 \nu_0}. 
\end{equation}

\subsection{Pseudospin Transfer Torques} 

In the following section we report on simulations in which 
we drive an interlayer current by keeping the top left and top right contacts grounded
and applying identical interlayer voltages, $V_{INT}$, at the bottom left and bottom right contacts. 

We choose this bias configuration so as to focus on interlayer currents that are relatively uniform.  
The transport properties depend only on the quasiparticle Hamiltonian and on the chemical potentials in the leads. Because the pseudospin effective field is the only term in the quasiparticle Hamiltonian which does not conserve the $\hat{z}$ component of the pseudospin, it follows that every quasiparticle wavefunction in the system must satisfy\cite{Nunez2006}
\begin{equation}
\label{Eq:torque}
\partial_t m_{ps}^{z} = - \vec{\nabla} \cdot \vec{j}^{z} - \frac{2}{\hbar} \, (\vec{m}_{ps} \times \vec{\Delta})_z =0
\end{equation}
where $\vec{j}^{z}$ is the $\hat{z}$ component of the pseudospin current contribution from that orbital, {\em i.e.} the difference between bottom and top layer number currents, and $\vec{m}_{ps}$ is the pseudospin magnetization of that orbital. For steady state transport, the quasiparticles satisfy time-independent Schr\"{o}edinger equations so that, 
summing over all quasiparticle orbitals we find,
\begin{equation}
\label{Eq:torque2}
2|\vec{m}_{ps}| | \vec{\Delta}|  \sin(\phi_{ps}-\phi_{\Delta}) =  2 t m_{ps}^{y} = \hbar \vec{\nabla} \cdot \vec{j}^{z}.
\end{equation}
In Eq. (\ref{Eq:torque2}), $\phi_{\Delta}$ is the planar orientation of $\vec{\Delta}$.
The first equality in Eq.(\ref{Eq:torque2}) follows from Eq.(\ref{Eq:deltadef}).
The pseudospin orbitals do not align with the effective field they experience
because they must precess between layers as they transverse the sample.  
The realignment of transport orbital pseudospin orientations alters the total pseudospin
and therefore the interaction contribution to $\vec{\Delta}$. 
The change in $\vec{m}_{ps} \times \vec{\Delta}$ due to transport currents is referred to 
here as the pseudospin transfer torque, in analogy with the terminology commonly found in metal spintronics.
Integrating Eq. (\ref{Eq:torque2}) across the sample from left to right and accounting for spin degeneracy, we find that 
\begin{equation} 
\frac{4 e t A  \langle m_{ps}^{y} \rangle}{\hbar} = I_{L}+I_{R}= I
\label{Eq:critcurrent} 
\end{equation} 
where $A$ is the 2D layer area, the angle brackets denote a spatial average, and 
$I_{L}$ and $I_{R}$ are the currents flowing from top to bottom at the left and right contacts.
(The pseudospin current $\vec{j}^{z}$ flows to the right on the right and is positive 
on the right side of the sample, but flows to the left and is negative on the left side of the sample.)
If the bias voltage can drive an interlayer current $I$ larger than $4 e t A  \langle m_{ps}^{y} \rangle/\hbar$,
it will no longer be possible to achieve a transport steady state.  Under these 
circumstances the interlayer current will oscillate in sign and the time-averaged current
will be strongly reduced.  In the next section we use a numerical simulation to assess 
the possibility of achieving currents of this size.   

A similar conclusion can be reached following a different line of argument.
The microscopic operator $\hat{J}$ describing net current flowing top layer($\uparrow$) to bottom layer($\downarrow$) given by\cite{Zheng1993, Park2006}
\begin{equation}\label{Eq:currentOp}
\begin{split}
\hat{J} &= \frac{-iet}{\hbar}\sum_{\bf{k},\bf{\sigma}}\left(c^\dagger_{\bf{k},\bf{\sigma},\uparrow}c_{\bf{k},\bf{\sigma},\downarrow}- c^\dagger_{\bf{k},\bf{\sigma},\downarrow}c_{\bf{k},\bf{\sigma},\uparrow}\right)\\
&=\frac{-2et}{\hbar}i\sum_{\bf{k}}\left(c^\dagger_{\bf{k},\uparrow}c_{\bf{k},\downarrow}- c^\dagger_{\bf{k},\downarrow}c_{\bf{k},\uparrow}\right)\\
&=\frac{4et}{\hbar}m_{ps}^y,
\end{split}
\end{equation}
where $\bf{k}$, $\bf{\sigma}$ are the momentum and spin indices respectively.
Here we have introduced a common notatition $\Delta_{SAS}=2t$ for the 
pseudospin splitting between symmetric and antisymmetric 
states in the absence of interactions and added a factor of $2$ to account for spin-degeneracy.
The pseudospin density operator $m_{ps}^{y}$ in Eq. (\ref{Eq:currentOp}) is given by
\begin{equation}
m_{ps}^y=\frac{-i}{2}\sum_{\bf{k}}\left(c^\dagger_{\bf{k},\uparrow}c_{\bf{k},\downarrow}- c^\dagger_{\bf{k},\downarrow}c_{\bf{k},\uparrow}\right).
\end{equation}
As a result, current density flowing in the given system may be written as
\begin{equation}\label{Eq:currentOp2}
\begin{split}
\left<\hat{J}\right> &= \frac{4et}{\hbar}\left<m_{ps}^y\right>\\
&= \frac{4et}{\hbar}\vert\vec{m}_{xy}\vert \sin{\phi}\\
&\leq \frac{4et}{\hbar}\vert\vec{m}_{xy}\vert =J_c,\\
\end{split}
\end{equation}
where $\vert \vec{m}_{xy}\vert=\sqrt{\langle m_x\rangle^2+\langle m_y\rangle^2}$, $\phi_{ps}$ is defined in Eq. (\ref{Eq:psfielddef}) and $J_c$ is defined critical current density. Assuming the current flow is uniform across the device, it is possible to obtain the critical current simply by multiplying Eq. (\ref{Eq:currentOp2}) by the system area $A$ to obtain
\begin{equation}\label{Eq:critI}
I_c=J_c \times A=\frac{4etA}{\hbar}\vert\vec{m}_{xy}\vert.
\end{equation}

\subsection{Simulation detail}  
Because we neglect disorder in this simulation
we may write the Hamiltonian in the form of decoupled 1D longitudinal channels
in the transport ($\hat{x}$) direction, taking proper account of 
the eigenenergy of transverse ($\hat{y}$) direction motion\cite{Venugopal2002}.
The calculation strategy follows a standard self-consistent field procedure. Given the density matrix of the 2D system, we can evaluate the mean-field quasiparticle Hamiltonian. For the interlayer tunneling part of the Hamiltonian we use the local-density approximation outlined in subsection A. The electrostatic potential in each layer is calculated from the charge density in each of the layers by solving a 2D Poisson equation using an alternating direction implicit method\cite{Stone1968} with appropriate boundary conditions. The boundary conditions employed in this situation were hard wall boundaries on the top and sides of the simulation domain and Neumann boundaries at the points where current is injected to insure charge neutrality\cite{Rahman2003}. Given the mean-field quasiparticle Hamiltonian and voltages in the leads, we can solve for the steady state density matrix of the two-dimensional bilayer using the NEGF method with a real-space basis.
The density-matrix obtained from the quantum transport calculation is updated at each state in the iteration process. 
The update density is then fed back into the Poisson solver and on-site potential is updated using the Broyden method\cite{Johnson1988} to accelerate self-consistency. The effective interlayer tunneling amplitudes are also updated and the loop proceeds until a desired level of self-consistency is achieved. The transport properties are calculated after self-consistency is achieved 
by applying the Landauer formula,\textit{ i.e.}  by using
\begin{equation}
I(V_{sd})=\frac{2e}{h}\int{T(E)[f_s(E)-f_d(E)]}
\end{equation}

\section{Numerical Results} 
\subsection{Linear response} 

\begin{figure}[t!] 
  \centering
    \includegraphics[width=0.45\textwidth]{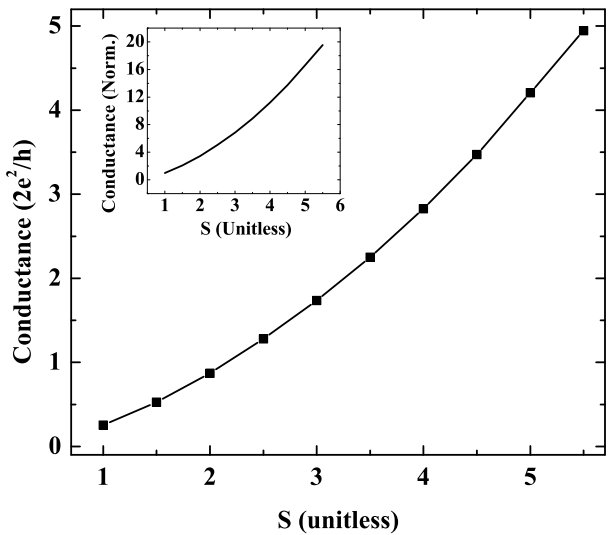}
  \caption{Plot of the height of the interlayer conductance as a function of the interlayer exchange enhancement S at interlayer bias $V_{T}-V_{B}=10~nV$. In the inset, we plot of the height of the interlayer conductance normalized to the non-interacting conductance (S=1) as a function of S with the same x axis. We clearly see that the height of the interlayer conductance follows an $S^2$ dependence. The single particle tunneling amplitude $t=\Delta_{SAS}/2=1~\mu eV$.} \label{FIG:Svaring}
\end{figure} 
\begin{figure}[t!] 
  \centering
    \includegraphics[width=0.5\textwidth]{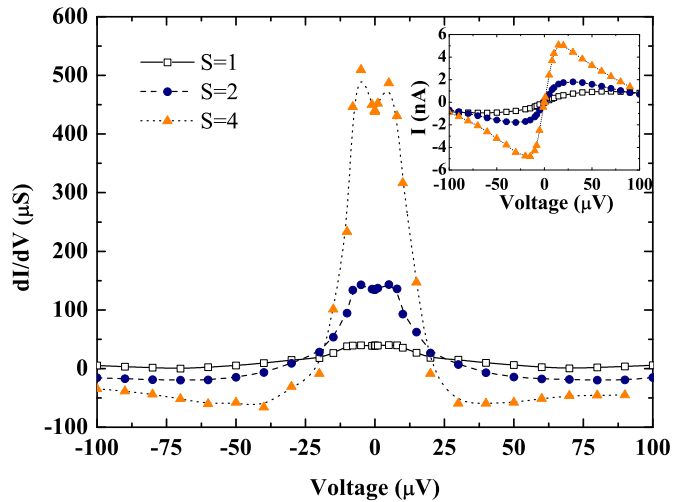}
  \caption{Plot of the differential conductance of the bilayer system at $S=4$ (orange triangle), $S=2$ (blue circle), and $S=1$ (opened rectangular) as a function of interlayer bias $V_B-V_T$ at 0K. In the inset, the same plot in negative interlayer bias with same y axis is plotted with x axis in log scale. In small bias window, almost constant interlayer transconductance is observed. After the interlayer bias reaches $V_{INT}\approx 10\mu V$, an abrupt drop in transconductance can be seen in the interacting electron cases of $S=2, 4$. We plot the current {\em vs.} bias voltage for each enhancement factor in the inset.} \label{FIG:TransG}
\end{figure} 
In Fig. (\ref{FIG:Svaring}), we plot the interlayer conductance at temperature $T=0$ as a function of the interlayer exchange enhancement, $S$ by setting $V_{B}=V_{INT}$  and $V_T=0$ in Fig. (\ref{FIG:Device}). 
The $S^2$ dependence demonstrates that the conductance in our nanodevice
is proportional to the square of the quasiparticle 
tunneling amplitude, as in bulk samples\cite{Misra2008, Eisenstein1991, Eisenstein19912, Zheng1993, Turner1996, Lyo2000}, and that the quasiparticle tunneling amplitude is approximately uniformly enhanced 
even though the finite size system is not perfectly uniform.
The range of $S$ used in this figure corresponds to the relatively modest enhancement factors that we expect
in bilayer systems in which the individual layers have the same sign of mass. 
For systems in which the quasiparticle masses are opposite in the two layers we expect values of $S$ that are significantly larger.
Spontaneous interlayer coherence, which occurs in a 
magnetic field\cite{Spielman2000, Min2008, Seamons2007, Zhang2008, Balatsky2004}
but is not expected in the absence of a field would be signaled by a divergence in $S$.

The physics of the results illustrated in Fig. (\ref{FIG:Svaring}) can be understood qualitatively by ignoring the finite-size-related spatial inhomogeneities present in our simulations and considering the simpler case in which 
there is a single bottom-layer source contact and a single top-layer drain contact on opposite 
ends of the transport channels.
The interlayer tunneling is then diagonal in transverse channel, and the transmission probability from top layer to bottom layer in channel $k$ is
\begin{equation}
\label{Eq:tintanal}
T_{interlayer}=\sin^2\left(\frac{\delta_kL}{2}\right).
\end{equation}
In Eq. (\ref{Eq:tintanal}), $\delta_k$ is the difference between the current direction wavevectors of the symmetric and antisymmetric states and $L$ is the system length in the transport direction. We may simplify the expression in Eq. (\ref{Eq:tintanal}) by rewriting it in terms of the Fermi velocity, $v_f$ in the transport-direction and making use of the small angle expansion of the \textit{sin} function to obtain
\begin{equation}
T_{interlayer}=\left(\frac{t_{eff}L}{\hbar v_f}\right)^2,
\end{equation}
where $t_{eff}$ is the quasiparticle tunneling amplitude proportional to the interlayer exchange enhancement $S$ resulting in the power law dependence seen in Fig. (\ref{FIG:Svaring}). As this approximate expression suggests, we find that transport at low interlayer bias voltages is dominated by the highest energy transverse channel, which has the smallest transport direction Fermi velocity.

\subsection{Transport beyond linear response} 

We have seen that at small bias voltages the inter-layer current is enhanced by interlayer exchange interactions.  
In Fig. (\ref{FIG:TransG}) , we compare our interlayer transport result with the $S=1$ case.  
We see that all of the curves are sharply peaked near zero bias, as in bulk 2D to 2D tunneling\cite{Misra2008, Eisenstein1991, Eisenstein19912, Zheng1993, Turner1996, Lyo2000}.
In all cases the decrease and change in sign of the differential conductance at higher bias voltages 
is due to the build-up of a Hartree potential 
difference ($\Delta_{z}$) between the layers which moves the bilayer away from its resonance condition.
The peak near zero bias is sharper for the enhanced interlayer tunneling 
(filled triangle and circle in Fig. (\ref{FIG:TransG})) cases.
To get a clearer picture of $S > 1$ transport properties in the non-linear regime,
we examine the influence of bias voltage on steady-state 
pseudospin configurations.  
Initially, all orbitals are aligned with the external pseudospin field, the inter-layer tunneling, and it 
follows from Eq. (\ref{Eq:critcurrent}) that $\phi_{ps}=0$.  The enhanced magnitude of this equilibrium 
pseudospin polarization is $m_0= \nu_0 S t = \nu_0 t_{eff}$.  When current flows between there 
must be a $\hat{y}$-component pseudospin, as we have explained previously, whose spatially averaged value 
is proportional to the current.  If it is possible to drive a current that is larger than allowed by 
Eq.(~\ref{Eq:critI}), it will no longer be possible to sustain a time-independent steady state.  

\begin{figure}[t!] 
\includegraphics[width=0.5\textwidth]{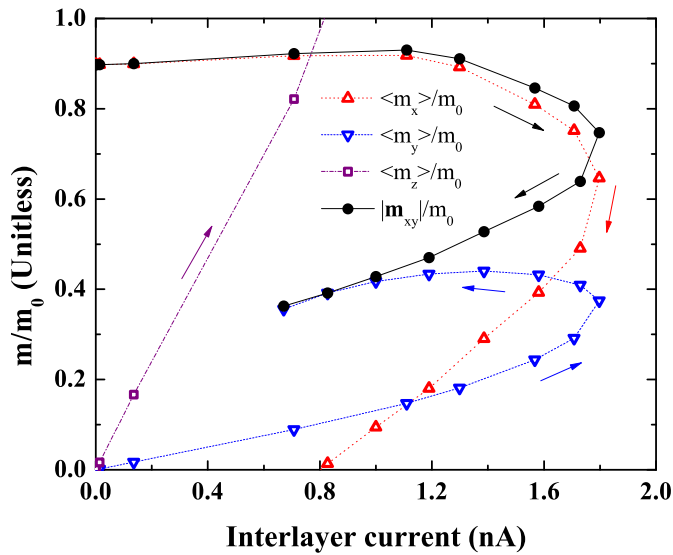}
\caption{Pseudospin polarization {\em vs.} current. The quantities are expressed in units of the 
value of $m_x$ at zero bias ($m_0$). The points in this plot correspond to the 
same points that appear in the inset of current {\em vs.} bias voltage in Fig.(\ref{FIG:TransG}).  For 
zero-current,  the pseudospin is in the $\hat{x}$ direction and has a value close to $m_0= 
\nu_0 S t$. $m_x$ decreases monotonically, $m_z$ increases monotonically, and $m_y$ increases nearly-monotonically before saturating at the largest field values. The arrow in the plot indicates the direction in which a bias voltage increases.} 
\label{FIG:mvsI}
\end{figure} 
\begin{figure}[t!] 
\includegraphics[width=0.5\textwidth]{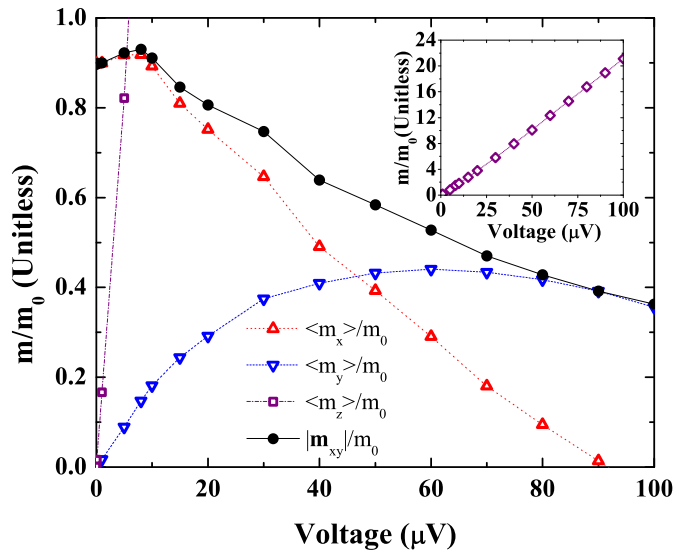}
\caption{Pseudospin polarization {\em vs.} bias voltage. The quantities are expressed in units of the 
value of $m_x$ at zero bias ($m_0$). The inset of this plot clearly shows that $m_z$ increases monotonically with bias voltage which eventually causes the current to decrease. The $\vert \vec{m}_{xy}\vert$ decreases only slowly with bias voltage because of the compensating effects of pseudospin rotation toward the $\hat{z}$ direction and an increase in the overall pseudospin polarization.
There is no steady state solution to the self-consistent transport equations beyond the largest bias voltage 
plotted here, because the pseudospin orientation $\phi_{ps}$ has reached $\pi/2$.} 
\label{FIG:mvsV}
\end{figure} 

In Figs.(\ref{FIG:mvsI}) and (\ref{FIG:mvsV}) we plot the magnitudes of the $\hat{x}$, $\hat{y}$, and $\hat{z}$ directed pseudospin fields, along with $|\vec{m}_{xy}|=\sqrt{m_x^2+m_y^2}$, evaluated at the 
center of the device, as a function of current and bias voltage, respectively.  The pseudospin densities are plotted in units of their equilibrium values.
The fifteen data points for each of these curves correspond to the 15 data points in the current {\em vs.} bias potential 
plot in Fig. (\ref{FIG:TransG}), so that the maximum current corresponds to a bias voltage of 
$\sim 20 {\rm \mu V}$ and the last data point corresponds to a bias voltage of $\sim 100 {\rm \mu V}$.   
The first thing to notice in this plot is that the $\hat{z}$ pseudospin component increases monotonically with 
bias voltage, as a potential difference between layers builds up.  This is the effect which eventually causes the 
current to begin to decrease.  The $\hat{x}$ component of pseudospin increases very slowly with bias voltage
in the regime where the current is increasing, but drops rapidly when current is decreasing.    
At the same time the $\hat{y}$ component of pseudospin evaluated at the device center rises steadily with current
until the maximum current is reached and then remains approximately constant.  When the two effects are combined
$|\vec{m}_{xy}|$ first increases slowly with bias voltage and then decreases slightly more rapidly.  
The relatively weak dependence of $|\vec{m}_{xy}|$ on current can be understood as a competition between two effects.
Because of the Coulomb potential which builds up and lifts the degeneracy between states 
localized in opposite layers, the direction of the pseudospin tilts toward the $\hat{z}$ direction, decreasing the 
$\hat{x}-\hat{y}$ plane component of each state.  At the same time the total magnitude of the pseudospin
field increases, increasing the pseudospin polarization.  In a uniform system the two effects cancel when 
a $\hat{z}$-direction pseudospin field is added to a uniform system.

\begin{figure}[t!] 
    \includegraphics[width=0.5\textwidth]{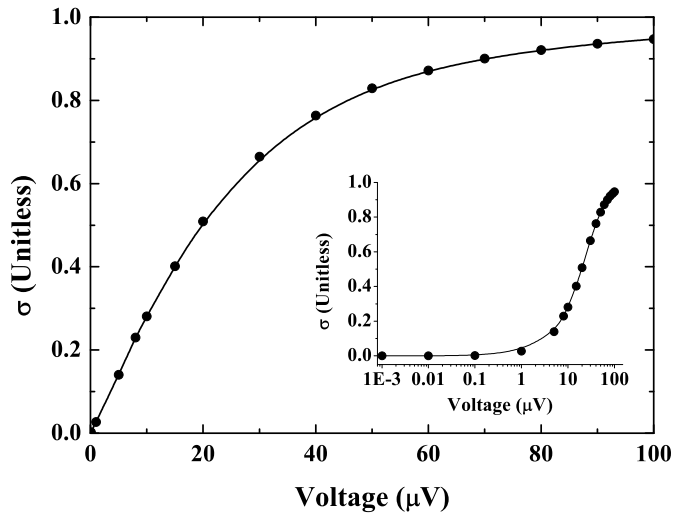}
  \caption{Quasiparticle layer-polarization parameter $\sigma$ as a function of interlayer bias. 
When the source-drain bias exceeds around $40~\mu V$,
the mismatch match between subband energy levels exceeds the 
width of the conductance peak, $\sigma$ increases rapidly, and the
$\hat{x}-\hat{y}$-plane pseudospin polarization and  
critical current decrease.  
The inset shows the same data in log scale to emphasize the low bias behavior.} \label{FIG:sigma}
\end{figure} 

Our simulations suggest the following scenario for how the critical current might be reached in bilayers.
The width of the linear response regime is limited by the lifetime of Bloch states which is set by 
disorder in bulk systems and in our finite-size system by the time for escape into the contacts.
In the linear-response regime the current is enhanced by a factor of $S^2$ by inter-layer 
exchange interactions.  The maximum current which can be supported in the steady-state 
is however proportional to the bare inter-layer tunneling and to the $\hat{x}-\hat{y}$ pseudospin 
polarization and is therefore enhanced only by a factor of $S$.  From this comparison we can 
conclude that more strongly enhanced inter-layer tunneling quasiparticle tunneling amplitudes 
(larger $S$) increases the chances of reaching a critical current beyond which the system 
current response is dynamic.  For the parameters of our simulation, the maximum current
of around $4 {\rm nA}$ is reached at a bias voltage of around $10 {\rm \mu V}$.  At this bias voltage 
the average $\hat{x}-\hat{y}$ plane angle of the pseudospin field is still less than $90^{\circ}$, 
and steady state response occurs.  At the bias voltage increases further the total current 
decreases, but the pseudospins become strongly polarized in the $\hat{z}$ direction. 
We illustrate this behavior in Fig.(\ref{FIG:sigma}) by plotting
\begin{equation}
\label{sigmadef}
\sigma=\Delta_z/\sqrt{\Delta_z^2+\Delta_{c}^2}
\end{equation}
{\em vs.} bias voltage, where $\Delta_z$ is $\hat{z}$ directional pseudospin field and $\Delta_{c}$ is a constant beyond which charge imbalance of the system becomes significant\cite{Bourassa2006}.
The magnitude of the $\hat{x}-\hat{y}$ direction pseudospin polarization consequently 
decreases.  In our simulations the critical current decreases more rapidly than the current 
in this regime.  As shown in Fig. (\ref{FIG:mvsI}), $\phi_{\rm ps}$ approaches $\pi/2$ 
($m_x \to 0$) at the largest bias voltages ($\sim 100 {\rm \mu V}$) for which we are able to 
obtain a steady-state solution of the non-equilibrium self-consistent equations.

\section{Discussions and Conclusions} 

For a balanced double quantum well system, the equilibrium electronic eigenstates have symmetric or antisymmetric 
bilayer wavefunctions.  When the bilayer is described using a pseudospin language, the difference between
symmetric and antisymmetric populations corresponds to a $\hat{x}$-direction pseudospin polarization
When the bilayer system is connected to reservoirs that drive current between layers, it is easy to show that 
the non-equilibrium pseudospin polarization must tilt toward the $\hat{y}$-direction.  The total 
current that flows between layers is in fact simply related to the total $\hat{y}$-direction pseudospin polarization.
These properties suggest the possibility that electron-electron interactions can qualitatively alter 
interlayer transport under some circumstances.  In equilibrium interactions enhance the
quasiparticle inter-layer 
tunneling amplitude, but not the total current that can be carried between layers for a given 
$\hat{y}$-direction pseudospin polarization.  If the inter-layer quasiparticle current can be driven to a 
value that is larger than can be supported by the inter-layer tunneling amplitude, 
the self-consistent equations for the transport steady state have no solution and time-dependent 
current response is expected.  This effect has not yet been observed, but is 
partially analogous to spin-transfer-torque oscillators in circuits containing magnetic metals.
We have demonstrated by explicit calculation for a model bilayer 
system that it is in principle to induce this dynamic instability in semiconductor bilayers.

The current-voltage relationship in semiconductor bilayers is characterized by a 
sharp peak in $dI/dV$ at small bias voltages, followed by a regime of 
negative differential conductance at larger bias voltages.  In our simulations the 
pseudospin instability occurs in the regime of negative differential conductance
where dynamic responses might also occur simply due to normal electrical instabilities.
It should be possible to distinguish these two effects experimentally, by varying the 
circuit resistance that is in series with the bilayer system.  The interaction effect might 
also be more easily realized in bilayers in which the resonant interlayer 
tunneling conductance is broadened by intentionally adding disorder.  

\begin{acknowledgements}
We acknowledge support for the Center for Scientific Computing from the CNSI, MRL: an NSF MRSEC (DMR-1121053) and NSF CNS-0960316 and Hewlett-Packard. 
YK is supported by Fulbright Science and Technology Award.
YK and MJG are supported by the ARO under contract number W911NF-09-1-0347. 
AHM was supported by the NRI SWAN program and by the Welch Foundation under grant TB-F1473. 
\end{acknowledgements}

\bibliographystyle{apsrev}
\bibliography{manuscript}		

\end{document}